\documentclass[twocolumn,showpacs,aps,groupedaddress,floatfix]{revtex4}

\usepackage{graphicx} 
\usepackage{dcolumn}
\usepackage{epsfig}
\usepackage{amssymb} 
\usepackage{amsmath} 
\usepackage{color} 

\renewcommand{\vr}{v_{\scriptscriptstyle 0}}
\newcommand{\thetaz}{\theta_{\scriptscriptstyle 0}}
\newcommand{\cz}{c_{\scriptscriptstyle 0}}
\newcommand{\pz}{p_{\scriptscriptstyle 0}}
\newcommand{\degre}{^{\circ}}

\newcommand{\Rr}{R_{\scriptscriptstyle 1}}
\newcommand{\Rs}{R_{\scriptscriptstyle 2}}
\newcommand{\pt}{p_{\scriptscriptstyle T}}
\newcommand{\pdt}{p_{\scriptscriptstyle T+\delta T}}
\newcommand{\acper}{t_{\hbox{\rm \tiny ac}}}
\newcommand{\prf}{f_{\hbox{\rm \tiny PRF}}}
\newcommand{\gammap}{\dot{\gamma}}

\newcommand{\smblacksquare}{{\scriptscriptstyle\blacksquare}}
\newcommand{\smtriangle}{{\scriptscriptstyle\triangle}}
\begin{document}

\title{High-frequency ultrasonic speckle velocimetry in sheared complex fluids}



\author{S\'ebastien Manneville}
\affiliation{Centre de Recherche Paul Pascal, Avenue Schweitzer, 33600 Pessac, FRANCE}
\author{Lydiane B\'ecu}
\affiliation{Centre de Recherche Paul Pascal, Avenue Schweitzer, 33600 Pessac, FRANCE}
\author{Annie Colin}
\affiliation{Centre de Recherche Paul Pascal, Avenue Schweitzer, 33600 Pessac, FRANCE} 

\date{\today}

\begin{abstract}
High-frequency ultrasonic pulses at 36~MHz are used to measure velocity
profiles in a complex fluid sheared in the Couette geometry. Our
technique is based on time-domain cross-correlation of ultrasonic
speckle signals backscattered by the moving medium.
Post-processing of acoustic data allows us to record a velocity profile
in 0.02--2~s with a spatial resolution of 40~$\mu$m over 1~mm.
After a careful calibration using a Newtonian suspension,
the technique is applied to a
sheared lyotropic lamellar phase
seeded with polystyrene spheres of diameter 3--10~$\mu$m. 
Time-averaged velocity profiles reveal the existence
of inhomogeneous flows, with both wall slip and shear bands,
in the vicinity of a shear-induced ``layering'' transition.
Slow transient regimes and/or temporal fluctuations
can also be resolved and exhibit complex
spatio-temporal flow behaviors with
sometimes more than two shear bands.
\end{abstract}





\maketitle

\section{Introduction}

Accurate measurements of flow velocities are essential for exploring
and understanding fluid dynamics. Such measurements raise important
technical problems whenever one needs them to be nonintrusive.
Indeed, many applications where direct access to the fluid is impossible 
perclude the use of local probes such as miniature piezoelectric pressure
probes, hot films or hot wires.
In other cases, the introduction of a probe inside the fluid may perturb
the flow or even the fluid structure itself,
and may lead to incorrect interpretations of the measurements. 
Thus, in general, much insight can be gained from some remote,
noninvasive sensing of the flow field.

In the following, we focus on the case of complex fluids under simple
shear.
A ``complex'' fluid may be characterized by the existence of a
``mesoscopic'' length located somewhere between the size of individual
molecules and the size of the sample \cite{Larson:1999}.
For instance, this intermediate
length scale is the particle diameter in a colloidal suspension,
the diameter of an oil droplet in an emulsion, or the radius of gyration
of polymer coils. The existence of such a supramolecular organization
(or ``microstructure'')
can lead to very complicated behaviors under flow and to inhomogeneous
velocity profiles under simple shear \cite{Larson:1999,Edimbourg:2000}.
Since the microstructure of a complex
fluid is very sensitive to local deformations, it is quite obvious that
a nonintrusive technique is required in order to measure velocity profiles.

The most
popular nonintrusive techniques, namely Particle Imaging Velocimetry (PIV)
or Laser Doppler Velocimetry (LDV), are based on the interaction
between light and seeding particles following the flow
\cite{Cummins:1977,Berne:1995}. However,
many complex fluids, such as emulsions, slurries,
or pastes, may not be transparent enough to allow the use of PIV or LDV.
Nuclear Magnetic Resonance (NMR) offers the possibility to image opaque
media \cite{Callaghan:1991,Hanlon:1998}
but requires the use of powerful magnets and remains expensive
and tricky to set up. On the other hand, ultrasound appears as an
efficient, cost-effective tool to measure velocity profiles in a large
range of fluids.

In this paper, we adapt a speckle-based ultrasonic velocimetry
technique \cite{Hein:1993,Jensen:1996} to complex fluid flows.
Although based on
the classical principle of backscattering by particles suspended in the
flow, our technique brings an original contribution to both the fields of
acoustical flow measurements and of complex fluid studies.
Indeed, by using high-frequency pulses
(frequencies larger than 20~MHz), we show that we
are able to measure velocity profiles in complex fluids sheared between
two plates separated by 1~mm with a spatial resolution of about 40~$\mu$m.
Depending on the required accuracy, a full velocity profile
can be obtained typically in 0.02~s to 2~s, which
makes it possible to resolve transient regimes or temporal fluctuations
of the flow.

The paper is organized as follows. We first explain in more details
why local velocity measurements are crucial to the
understanding of complex fluids.
Examples are given that show that inhomogeneous flows may occur even
in simple shear geometries and at small imposed stresses. The third section
is devoted to a brief review of existing ultrasonic techniques for
measuring flow velocities. We then present the electronic setup
and the data analysis used for high-frequency ultrasonic
speckle velocimetry (USV). Section~\ref{s.calib} deals with the
calibration step necessary to obtain quantitative estimates of the
velocity. In Section~\ref{s.lamellar}, high-frequency
USV is applied to a particular complex fluid: an aqueous solution
of surfactant. The ultrasonic data reveal the existence
of inhomogeneous flows, with both wall slip and shear bands, as well as complex
spatio-temporal behaviors during transient regimes.
Finally, in light of the present results, the technique is compared
to other nonintrusive tools used in the field of complex fluid flows.
We emphasize on the promising spatio-temporal resolution of USV and
discuss its possible future applications.

\section{Shear flow of complex fluids: why are local measurements essential?}
\label{s.local}

\subsection{Shearing a Newtonian fluid in the Couette geometry}

\begin{figure}[htbp]
\begin{center}
\scalebox{0.85}{\includegraphics{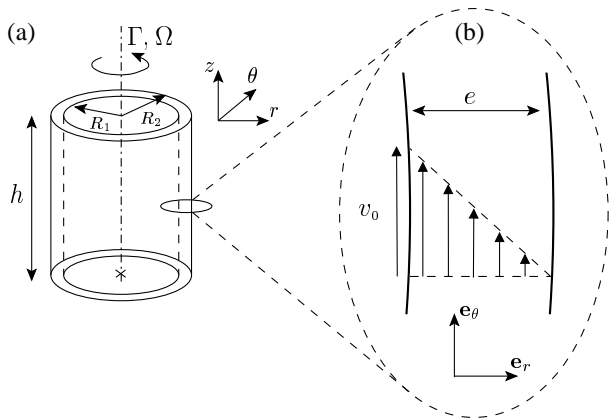}}
\end{center}
\caption{\label{f.couette}(a) Sketch of the Couette cell used in our
experiments. (b) Enlargement of the shear flow inside the
gap between the two cylinders in the case of a Newtonian fluid below
the onset of the Taylor-Couette instability.}
\end{figure}

It is well-known that shear flows of simple fluids are laminar at low
velocities and may become unstable above a critical Reynolds number due
to hydrodynamic instabilities \cite{Tritton:1988}.
A classical shear geometry is
the {\it Couette geometry} in which the fluid is sheared between two
concentric cylinders as sketched on Fig.~\ref{f.couette}. Usually
the inner cylinder of radius $\Rr$ is rotating while the outer cylinder
of radius $\Rs$ remains fixed. The two cylinders are thus called
``rotor'' and ``stator'' respectively. The distance $e=\Rs-\Rr$
between the two walls is called the ``gap'' of the Couette cell.
This Couette geometry will be used throughout the paper
as well as the cylindrical coordinates
$(r,\theta,z)$. We will denote
$x=r-\Rr$ the distance from the inner cylinder and
$\mathbf{v}=(v_r,v_\theta,v_z)$ the velocity vector. 

\begin{figure}[htbp]
\begin{center}
\scalebox{0.85}{\includegraphics{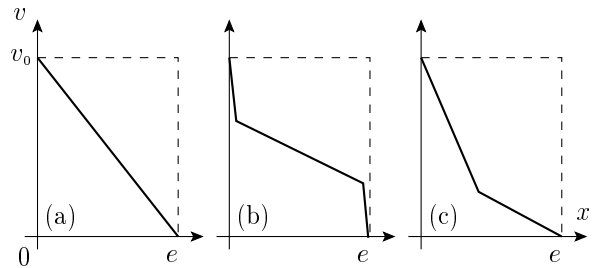}}
\end{center}
\caption{\label{f.exs}Examples of orthoradial
velocity profiles observed in a small gap
Couette geometry. (a) Newtonian fluid sheared
below the onset of the Taylor-Couette instability. (b) Wall slip
in a complex fluid: apparent slippage results from the presence
of thin lubricating layers near the walls. (c) Shear banding
in a complex fluid near a shear-induced transition: two differently
sheared bands coexist in the gap of the cell.}
\end{figure}

A Couette device can work in two different modes: either a torque
$\Gamma$ is imposed on the rotor and the rotor angular velocity $\Omega$
is measured, or $\Omega$ is imposed and the torque $\Gamma$ is measured.
For a Newtonian fluid
at small velocities, the flow is stationary
and purely orthoradial $\mathbf{v}=(0,v_\theta(r),0)$.
The tangential velocity $v(x)\equiv v_\theta(r)$
decreases from the rotor velocity $\vr=\Rr\Omega$
down to zero on the outer wall.
In the small gap approximation {\it i.e.} when $e\ll\Rr$, the
velocity profile is linear as shown
in Fig.~\ref{f.exs}(a). The shear rate
$\gammap$ ($r\theta$-component of the rate-of-strain tensor)
and the shear stress $\sigma$ (tangential force per unit surface
{\it i.e.} $r\theta$-component of the stress tensor)
are then almost uniform across the gap and are simply given by:
\begin{eqnarray}
\gammap\equiv r\frac{\partial (v_\theta/r)}{\partial r}
&=& \frac{\Rr\Omega}{e}=\frac{\vr}{e}\, ,\label{e.gammarheo}\\
\sigma &=& \frac{\Gamma}{2\pi\Rr H}\, ,
\label{e.sigmarheo}
\end{eqnarray}
where $H$ is the height of the Couette cell.
The viscosity $\eta$ of the sample can thus be computed from the
rheological data $\gammap$ and $\sigma$ by $\eta=\sigma/\gammap$,
as long as the flow remains laminar and stationary.

\subsection{Inhomogeneous flows in sheared complex fluids}
\label{ss.inhom}

\subsubsection{Apparent wall slip}

In complex fluids, the picture of the Couette flow may be radically different.
For instance,
when a complex fluid is confined between two plates and depending
on the roughness properties of the plates, the fluid velocity close
to the walls may strongly differ from that of the walls (see
Fig.~\ref{f.exs}(b)): the fluid
{\it slips}. Slippage is usually explained by the existence of two
thin lubricating layers in which the fluid structure is very different
from that in the bulk. For instance, in a colloidal suspension,
electrostatic or steric effects next to the walls may induce a
depletion of the particles on a small
distance, leaving a much less viscous fluid film at the walls and leading
to this apparent wall slip \cite{Barnes:1995}. In concentrated
oil-in-water emulsions, wall slip is due to the presence of very thin
water layers close to the walls.

Even if the bulk fluid remains Newtonian,
the velocity profile between the two plates may be quite different
from that expected in a uniformly sheared fluid that does not slip
(see Fig.~\ref{f.exs}(a)). Wall slip in complex fluids
raises important industrial problems and is often difficult to detect
and to assess quantitatively. In particular, computing the viscosity
from Eqs.~(\ref{e.gammarheo}) and (\ref{e.sigmarheo}) can be very
misleading since the effective shear rate in the bulk is
sometimes orders of magnitude smaller than $v_0/e$ \cite{Barnes:1995}.

\subsubsection{Hydrodynamic and elastic instabilities}

Moreover, even in the absence of wall slip and even for Newtonian fluids,
above a critical rotation speed of the rotor, the laminar flow becomes unstable
and vortices develop due to the centrifugal force
({\it Taylor-Couette instability}). Upon
further increasing the rotation speed, this three-dimensional,
inhomogeneous but still stationary
flow turns into a time-dependent turbulent flow
\cite{Tritton:1988}. The
procedure based on Eqs.~(\ref{e.gammarheo}) and (\ref{e.sigmarheo})
for measuring $\gammap$ and $\sigma$ and computing the
viscosity then completely fails because the
flow is no longer laminar and stationary.

About fifteen years ago, complex fluids such as
entangled polymer solutions were shown to present inhomogeneous flows
even when sheared at rotor velocities
far below the onset of the Taylor-Couette instability.
These unstable flows result from purely
{\it elastic instabilities} \cite{Larson:1990,Groisman:1998}.
Although due to the elasticity of the fluid
rather than visco-inertial effects, the phenomenology of such
instabilities is very similar to classical hydrodynamic instabilities
but occur at very low Reynolds numbers.

\subsubsection{Shear-induced structures and shear-banded flows}

Beside those inertial and elastic instabilities,
more subtle effects may come into play that are due to the very nature
of complex fluids. Indeed, application of shear to a complex fluid
may completely change its microstructure leading to a new
{\it shear-induced structure}. Such a phenomenon may be
called a {\it shear-induced transition} or a
{\it material instability} in the sense that it results from
a strong coupling between the shear flow and the microstructure of the fluid
\cite{Goddard:2003}.

A material instability usually shows up on rheological data as
a sudden jump in the viscosity when shear is increased.
For instance, if the shear-induced structure is much less viscous
than the complex fluid at low shear, then a large drop of the visocity
may be observed and the transition is called a shear-thinning transition.
During such a transition, the fluid is assumed to phase-separate into
two states: a shear-induced state that flows at a given shear rate and
coexists with the old structure, thus giving rise to two shear bands
\cite{Spenley:1993,Spenley:1996,Porte:1997}.
The corresponding {\it shear-banded} velocity profile is sketched in
Fig.~\ref{f.exs}(c).

In order to investigate those transitions, most previous studies
have focused on global measurements such as the viscosity of the sample as
a function of time \cite{Wheeler:1998,Bandyopadhyay:2001,Wunenburger:2001}.
Indeed, the time series $\eta(t)$ sometimes display
complex temporal fluctuations on time scales
characteristic of the microstructure rearrangements. However, such global
measurements do not provide any information on the spatial structure
of the flow \cite{Salmon:2002}. Thus, other studies have focused
on the local characterization of the fluid structure using birefringence
neutron, x-ray or light scattering
and showed the coexistence of bands of different microstructures
in the vicinity of shear-induced transitions
\cite{Schmitt:1994,Cappelaere:1997,Eiser:2000}.

As far as the flow field is concerned, rather
few local velocity measurements
are reported in the complex fluid literature \cite{Edimbourg:2000,Mueth:2000}.
The presence of inhomogeneous flows in some surfactant mixtures
(``wormlike micelles'') was first unveiled using NMR
\cite{Mair:1996,Britton:1999}.
However, the existence of shear bands in the velocity profiles
was firmly ascertained only recently using dynamic
light scattering (DLS) in heterodyne mode, a technique close to LDV \cite{Salmon:2003b,Salmon:2003c}.

From the various examples cited above, it
is quite clear that local measurements
are essential, for instance to
precisely follow the shear bands in both
space and time. In general, a velocity profile with about 20 points
accross a 1~mm gap, {\it i.e.} with a resolution of 50~$\mu$m,
will carry enough spatial information on the flow to allow
quantitative conclusions on possible inhomogeneous flow profiles.

\section{Measuring fluid flows with ultrasound: a short review}
\label{s.review}

\subsection{Continuous wave systems}

Contrary to electromagnetic waves, the {\it phase} of acoustic
waves is easily accessible using piezoelectric transducers.
Moreover, ultrasound may propagate deeply into optically
opaque media. These two properties have led to the spectacular
development of ultrasonic imaging techniques since the early 1950s
particularly in the{ \it biomedical} domain.

When an ultrasonic wave travels through biological tissues,
it gets scattered by density and/or compressibility inhomogeneities
 \cite{Pierce:1994}.
If the scatterer moves in a flow, like for instance red blood cells
inside blood vessels, its motion induces a
Doppler shift in the wave frequency.
Early systems for measuring blood velocities
were based on the estimation of this Doppler frequency shift
using a monochromatic ultrasonic wave \cite{Jensen:1996}.

Unfortunately, even if these ``true Doppler'' systems helped to detect
the occlusion of blood vessels,
they did not offer any spatial resolution.
Indeed, if one uses a continuous wave, one gets a very
good resolution on the frequency shift (and thus on the velocity) but
one also loses any temporal information on the echo arrival time (and
thus on the scatterer position, as explained below).
Continuous Doppler systems may thus be very
efficient in the case of a single scatterer suspended
in a flow whose position is measured independently
\cite{Mordant:2002}.

\subsection{Pulsed wave systems}

In order to discriminate between different scatterers
in space and to measure velocity profiles, a solution is
to use short acoustic {\it pulses}.
Indeed, if an ultrasonic pulse is sent by a transducer
through a scattering medium, a series of backscattered echoes
can be recorded on the same transducer. The precise nature
of the backscattered (BS) signal will be discussed at length
in Sec.~\ref{s.usv}. At this point, it is only important to note
that the arrival times $t_k$ of the echoes
are directly linked to the position $y_k$ 
of the scatterer along the acoustic beam by $y_k=\cz t_k/2$,
where $\cz$ is the effective sound speed and the factor 2 accounts
for the round trip from the transducer to the
$k^{\hbox{\tiny th}}$ inhomogeneity and back to the transducer.

The idea is then to follow the scatterer motion
through the evolution of the BS signals: 
between successive pulses, echoes will move along with the flow
and their positions may be tracked in time.
Although the Doppler effect is not used in pulsed wave systems,
these are still mistermed ``pulsed Doppler systems'' in the literature
and the technique is referred to as ``ultrasonic Doppler velocimetry'' (UDV)
or ``ultrasonic velocity profiler'' (UVP) \cite{Takeda:1995}. 
This confusion may be due to the fact that, 
owing to the constraint of real-time display in commercial devices,
BS signals are sampled only at a few given depths and that these
sampled signals are analyzed in the frequency domain
\cite{Jensen:1996}.

UDV has been successfully applied
to various classical problems in hydrodynamics such as the Taylor-Couette
instability \cite{Takeda:1999},
cylinder wakes \cite{Peschard:1999}, or
magnetic fluid flows \cite{Brito:2001}.
The fluid has to be seeded with small particles
that scatter ultrasound. Such acoustic ``contrast agents'' are assumed
to follow the flow as Lagrangian tracers. The UVP has also been
used fairly recently for in-line
rheological studies of concentrated suspensions
in pipe flows \cite{Ouriev:2002}.

If one drops the constraint of real-time display,
the whole BS signals may be stored
for post-processing. Direct tracking of
the BS signals in the time domain
is the subject of recent technical developments such as
ultrasonic speckle velocimetry which will be addressed in the
next Section.



\section{High-frequency ultrasonic speckle velocimetry}
\label{s.usv}

\subsection{Speckle tracking techniques}

When the moving medium contains a lot of scatterers per
unit volume, the BS signal results from the interferences
of all the backscattered waves and appears as a complex,
high-frequency signal called {\it ultrasonic speckle}
(see Fig.~\ref{f.signal} for a typical speckle signal and
Sec.~\ref{s.analysis} for more details, in particular about
the possibility of multiple scattering).
The development of fast digitizers with large memories
has allowed the full recording of backscattered BS signals
for post-processing. Various time-domain algorithms for tracking
the motion of ultrasonic echoes have been proposed
\cite{Bonnefous:1986,Foster:1990,Ferrara:1994,KaisarAlam:1995,Bohs:1995}.
So far, such {\it speckle tracking} techniques have mainly been applied
to the measurement of tissue motion or
blood flow \cite{Hein:1993,Jensen:1996}.
They have also been adapted to the UVP system and tested on pipe
flows to provide high
time resolution (of about 0.5~ms per profile) \cite{Ozaki:2002}.
Recently, one of us introduced
a combination of a 2D echographic system and speckle tracking
that we called 2D ultrasonic speckle velocimetry (USV)
\cite{Sandrin:2001}.
This system allowed us to image strong vortical flows in Newtonian
fluids in two dimensions \cite{Manneville:2001b}.

Here we propose to adapt such a technique to complex fluids confined
in a Couette cell of gap $e\simeq 1$~mm. As explained in
Section~\ref{ss.inhom}, one needs to measure velocity profiles with a
spatial resolution of typically 50~$\mu$m.
However, frequencies commonly used in commercial ultrasonic velocimeters
are in the range $f=1$--10~MHz,
and with a typical sound speed $\cz\simeq 1500$~m.s$^{-1}$, this yields
a resolution of about $\lambda=\cz /f\simeq 0.15$--1.5~mm.
Thus, conventional ultrasound does not permit very
fine measurements and one has to turn to utrasound
with frequencies larger than 20~MHz. Again, high frequencies have
first attracted the biomedical community. High-resolution
ultrasonic images have been recently obtained in dermatology,
ophtalmology and stomatology \cite{Turnbull:1995,Berson:1999}
or for measuring blood flow in the microcirculation
\cite{Ferrara:1996,Christopher:1997}. To our
knowledge, no such high-frequency measurements (above 20~MHz)
have been reported
so far in classical hydrodynamics nor in the field of complex fluids.

Note that the resolution cannot be increased indefinitely by increasing
the frequency because {\it attenuation} of sound waves sharply increases
with frequency (as $f^2$ in water). Attenuation is due to the combination
of thermo-viscous absorption and scattering. The
signal-to-noise ratio rapidly deteriorates as $f$ increases.
We found that for our application to complex fluids,
$f=36$~MHz realizes a good compromise between resolution and attenuation
over propagation distances of a couple of centimeters.
Finally, since transducer arrays are not
available at such a high frequency, the technique will have to be 
restricted to one-dimensional measurements.

\subsection{High-frequency USV electronic system}

\begin{figure}[htbp]
\begin{center}
\scalebox{1}{\includegraphics{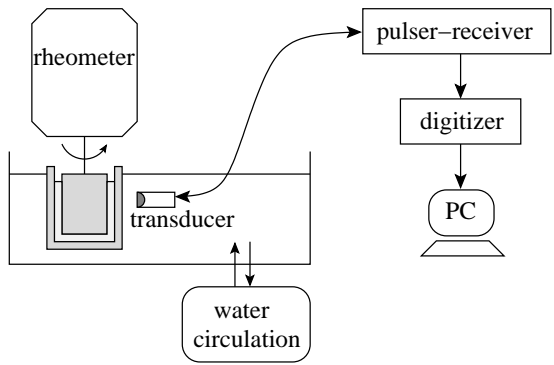}}
\end{center}
\caption{\label{f.system}Electronic system used for high-frequency USV.}
\end{figure}

Figure \ref{f.system} presents our USV electronic system.
Focused ultrasonic pulses are generated by a PVDF
piezo-polymer immersion transducer of
central frequency $f=36$~MHz (Panametrics PI 50-2).
The focal distance is 11.6~mm and the active element diameter
is 6.3~mm. The transducer bandwidth is 11 MHz at -3~dB.
The axial and lateral resolutions given by the manufacturer at -3~dB
are 30~$\mu$m and 65~$\mu$m respectively and the depth of field is
about 1~mm.

The transducer is controlled by a pulser-receiver unit (Panametrics 5900PR).
The pulser generates 220~V~pulses with a rise
time of about 1~ns. The pulse
repetition frequency (PRF) is tunable from 0 to 20~kHz.  The receiver is
equipped with a 200~MHz broadband amplifier of maximum voltage gain
54~dB as well as a set of selectable high-pass and low-pass filters.
BS signals are sampled at $f_s=500$~MHz, stored on a high-speed
PCI digitizer with 8 Mb on-board memory (Acqiris DP235), and later
transferred to the host computer for post-processing.

\subsection{Experimental setup in the Couette geometry}

The shear flow is generated in a Plexiglas Couette cell
with $\Rr=24$~mm, $\Rs=25.07$~mm,
and $H=30$~mm. The rotation of the inner cylinder is controlled
by a standard rheometer (TA Instruments AR1000). The whole cell is
surrounded by water whose temperature is kept constant to within
$\pm 0.1\degre$C. The thickness of the stator is 2~mm everywhere
except for a small rectangular window where the minimal thickness
is 0.5~mm in order to avoid additional attenuation due to the Plexiglas.

\begin{figure}[htbp]
\begin{center}
\scalebox{0.9}{\includegraphics{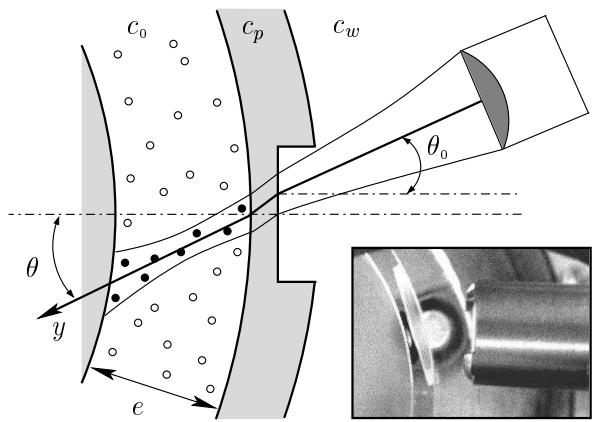}}
\end{center}
\caption{\label{f.setup}Experimental setup for USV in complex fluids
sheared in the Couette geometry. $\cz$, $c_p$, and $c_w$ stand
for the speed of sound in the complex fluid, in Plexiglas, and in water
respectively. The inset shows a picture of the transducer and
the stator as seen from above once the rotor has been removed.}
\end{figure}

Ultrasonic pulses are incident on the stator with a given angle
$\thetaz\simeq 20\degre$ relative to
the normal to the window in the stator as sketched in Fig. \ref{f.setup}.
They travel through Plexiglas and enter the gap with an angle $\theta$
that is given by the law of refraction. Since the sound
speed of the working fluid may differ from that of water,
and since the precise value of $\theta$ depends on the
exact arrangement of the acoustic beam relative the stator,
$\theta$ will be taken as an unknown until a careful
calibration procedure is completed (see Section~\ref{s.calib}).

Once inside the fluid, ultrasonic pulses get scattered
by inhomogeneities that can be
either naturally present (oil droplets in an emulsion
for instance) or artificially introduced to enhance the acoustic
contrast. The total round trip for a pulse travelling from
the transducer to the rotor and back to the transducer lasts about
15~$\mu$s.

The position of the transducer is tuned so that the 1~mm gap
lies into the focal spot in order to optimize the signal-to-noise
ratio. Typical recorded
BS signals are 1000 point long, which
corresponds to a transit time of 2~$\mu$s or
equivalently to a depth of roughly 1.5~mm inside the medium.
Allowing for a fraction of a millimeter before the stator and after
the rotor helps to locate the two walls. Note that spurious 
reflections on water--Plexiglas or Plexiglas--fluid interfaces
are minimized by carefully choosing the angle $\thetaz$.
In any case, even in the absence of interfaces, such an angle is
necessary to get a non-zero projection of the velocity vector
along the acoustic axis as explained below.

\subsection{Data analysis involved in USV}
\label{s.analysis}

\subsubsection{Basic principle of USV}

\begin{figure}[htbp]
\begin{center}
\scalebox{1}{\includegraphics{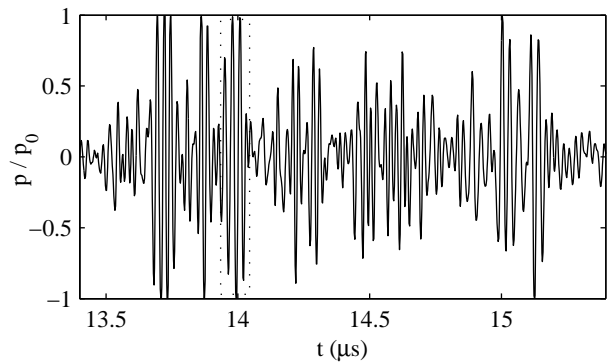}}
\end{center}
\caption{\label{f.signal}Speckle signal recorded in a
1~\% wt. suspension of polystyrene spheres. The voltage
$V(t)$ from the transducer output is directly proportional
to the pressure $p(t)$. The signal was normalized by its
maximum value $V_{\scriptscriptstyle 0}$, which yields
$V/V_{\scriptscriptstyle 0}=p/\pz$ where $\pz$ is the
maximum pressure. The sampling
frequency is $f_s=500$~MHz and the recorded signal is
1000 point long. The rotor velocity is $\vr=23.5$~mm.s$^{-1}$.
The dotted rectangle denotes the time window used in
Figs.~\ref{f.signal_zoom} and \ref{f.correl}.}
\end{figure}

Figure~\ref{f.signal} shows a typical signal backscattered 
by a dilute suspension of polystyrene spheres
and normalized by the maximum amplitude. The noise level 
is about 10~mV in all our experiments and typical
signal amplitudes are in the range 50--200~mV corresponding
to a signal-to-noise ratio of 10--30~dB.

Under the assumption of {\it single scattering}, the
speckle signal received between times $t$
and $t+\Delta t$ can be interpreted as interferences
coming from scatterers located between $y$ and
$y+\Delta y$, where $y=\cz t/2$, $\Delta y=\cz\Delta t/2$, and $y$ is the
distance from the transducer along the acoustic beam. The approximation
that no multiple scattering takes place is thus essential in order to
use the echographic conversion rule $y=\cz t/2$.
Multiple scattering is avoided by carefully controlling the amount
and the properties of the scatterers (see Sec.~\ref{s.calib}).

\begin{figure}[htbp]
\begin{center}
\scalebox{1}{\includegraphics{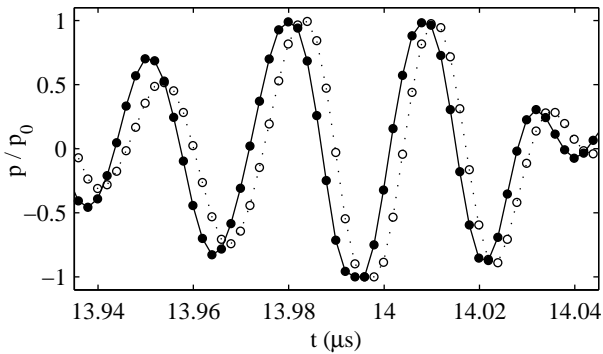}}
\end{center}
\caption{\label{f.signal_zoom}Zoom over 4 acoustic periods of
the speckle signal $p/\pz$. Bullets
($\bullet$) correspond to the speckle signal $\pt$ shown in Fig.~\ref{f.signal}
received after a pulse is sent at time $T$.
Open circles ($\circ$) correspond to the signal $\pdt$ recorded
over the same time window after a second pulse is sent 1~ms later at
$T+\delta T$ and when shear is applied. The motion of the
scatterers result in a shift to the right of the signal
{\it i.e.} the scatterers move away from the transducer.
The rotor velocity is $\vr=23.5$~mm.s$^{-1}$.}
\end{figure}

When the fluid is submitted to a shear flow, the interference 
pattern changes as the scatterers move along. Figure~\ref{f.signal_zoom}
shows two signals corresponding to two pulses separated by $\delta T=1$~ms
and plotted over 4 acoustic periods. A global shift
to the right of $\delta t\simeq 3$~ns is clearly visible. With
$\cz\simeq 1500$~m.s$^{-1}$, this corresponds to
$\delta y=\cz\delta t/2\simeq 2~\mu$m
and to a velocity $v_y=\delta y/\delta T \simeq 2$~mm.s$^{-1}$ projected
along the acoustic axis.

\subsubsection{Cross-correlation algorithm}

In order to get an accurate estimate of the time-shift $\delta t$,
we use the simple cross-correlation algorithm described below.
Two consecutive signals $\pt$ and $\pdt$ recorded after
pulses sent at times $T$ and $T+\delta T$ are
cross-correlated over small time windows of width
$\Delta t=4\acper\simeq 0.11~\mu$s,
where $\acper=1/f$ is the acoustic period. More precisely, the
following cross-correlation coefficient is computed according to:
\begin{equation}
C_k(\tau)=\sum_{t'=t_k-\Delta t/2}^{t_k+\Delta t/2}
\,\pt(t')\,\pdt(t'+\tau)\, ,
\label{e.crosscorr}
\end{equation}
where the $k^{\hbox{\rm \tiny th}}$ time window is centered
around $t_k=t_s+k\Delta t/2$ and $t_s\simeq 13.5~\mu$s is a
reference time that corresponds to the beginning of the gap.
$t_s$ depends on experimental parameters such as
the working temperature and on the exact arrangement of the
transducer relative to the Couette cell
(see Sec.~\ref{s.calib}). The times $t_k$ correspond to the centers of
the various time windows over which the signals are split. They will be
later converted into positions $y_k$ at which the velocity is measured using
$y_k=\cz (t_k-t_s)/2$. Such
windows correspond to slices of width $\Delta y=2\lambda\simeq 80~\mu$m 
along the acoustic axis separated by $\Delta y/2\simeq 40~\mu$m.

\begin{figure}[htbp]
\begin{center}
\scalebox{1}{\includegraphics{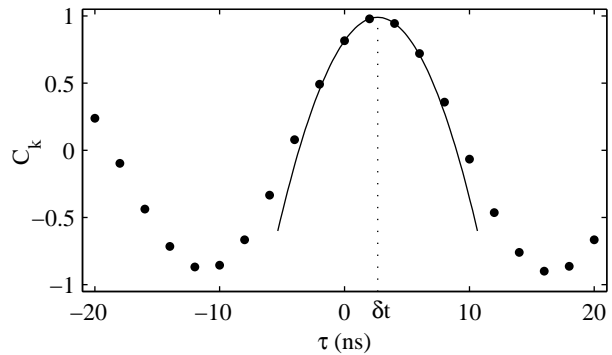}}
\end{center}
\caption{\label{f.correl}Correlation function $C_k(\tau)$ obtained by
using Eq.~(\ref{e.crosscorr}) on the two signals of Fig.~\ref{f.signal_zoom}
($\bullet$). The solid line corresponds to the parabolic
interpolation $\tilde{C}_k(\tau)$
and the dotted line indicates the position $\delta t\simeq 2.7$~ns of
the maximum of $\tilde{C}_k$.} 
\end{figure}

As shown in Fig.~\ref{f.correl},
the correlation function $C_k(\tau)$ is then interpolated around its
main maximum by a parabola and the value $\delta t$ of $\tau$ that
maximizes this parabola is extracted:
\begin{equation}
\tilde{C}_k(\delta t)=\max_{\tau} \left(\tilde{C}_k(\tau)\right)\, ,
\label{e.dt}
\end{equation}
where $\tilde{C}_k$ denotes the parabolic interpolation of $C_k$.
Note that $\delta t$ depends both on the position $t_k$ of the
time window (fast ``ultrasonic time'') and on the time $T$ for
which the cross-correlation is performed (slow ``pulse time'').
Thus $\delta t$ will be noted as a function of the two variables
$t_k$ and $T$ below.

Finally, it is straightforward to convert time-shifts $\delta t$
into velocities using:
\begin{equation}
v_y(y_k,T) = \frac{\cz}{2}\,\,\frac{\delta t(t_k,T)}{\delta T}\, ,
\label{e.vy}
\end{equation}
where
\begin{equation}
y_k=\cz (t_k-t_s)/2
\label{e.yk}
\end{equation}
is the position of the center of the window
along the acoustic axis. The origin $y=0$ is taken at the
stator--fluid interface. Equation~(\ref{e.vy}) thus yields the
projection $v_y$ of the velocity vector along the acoustic axis
at the discrete positions $y_k$ and at time $T$.

\subsubsection{Limits of the technique}

The procedure based on the interpolation of $C_k(\tau)$
allows us to measure time-shifts as small as
$\delta t\simeq 0.2$~ns that correspond to displacements
$\delta y\simeq 0.15~\mu$m.
Since the PRF $\prf=1/\delta T$ can be made as small as desired, 
infinitesimal velocities could in principle be measured. However,
at very small shear rates, decorrelation of the BS signals
may occur due to Brownian motion of the scatterers or to low-frequency
mechanical vibrations in the experimental setup.
Consequently, we will
avoid using PRFs smaller than 10~Hz and we estimate
the minimum measurable velocity at about 1~$\mu$m.s$^{-1}$.

On the other hand, the maximum measurable velocity is reached when
$\delta y\simeq\lambda\simeq 40~\mu$m. Indeed,
displacements greater than the
acoustic wavelength cannot be measured using
the cross-correlation algorithm described above unless
phase unwrapping and continuity conditions are implemented,
which will not be considered here \cite{Manneville:2001b}.
With the highest achievable PRF $\prf=20$~kHz, a
40~$\mu$m maximum displacement means that velocities above
0.8~m.s$^{-1}$ may not be accessed using USV. Note however
that this limiting range of [1~$\mu$m.s$^{-1}$ -- 0.8~m.s$^{-1}$]
applies to the velocity $v_y$ projected along the acoustic axis.
This range may be extended by varying the incidence angle $\theta$
(see Sec.~\ref{s.conversion} and Eqs. (\ref{e.v})--(\ref{e.x}) below).

Finally, Eq.~(\ref{e.vy}) shows that a velocity profile could in principle
be obtained by cross-correlating only two successive BS signals {\it i.e.}
within a time interval $\delta T$ that can be made very small depending
of the PRF. In practice, as seen on Fig.~\ref{f.signal},
the speckle amplitude is never uniform. Locally, destructive interferences
or the absence of scatterers may lead to signal levels too small to
be analyzed (see the signal of Fig.~\ref{f.signal} around $t\simeq 14.1$
or $14.4~\mu$s for instance). Time windows where the signal amplitude
does not reach a given threshold (typically 20~\% of the maximum amplitude)
are thus left out of the analysis. To recover a full velocity profile,
some averaging is then performed over several sucessive cross-correlations:
\begin{equation}
v_y(y_k) = \left<v_y(y_k,T)\right>_T\, .
\label{e.vymoy}
\end{equation}

\begin{figure}[htbp]
\begin{center}
\scalebox{1}{\includegraphics{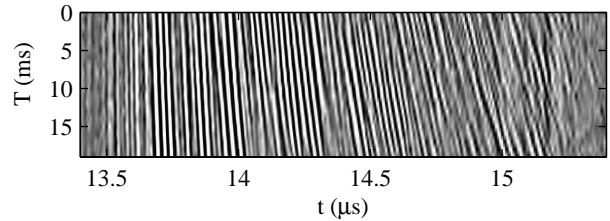}}
\end{center}
\caption{\label{f.data}BS signals corresponding to 20 pulses sent
every $\delta T=1$~ms. The normalized pressure $p/\pz$
is coded in gray levels. In order to highlight the areas with
strong speckle signal,
the colormap was saturated so that white (black resp.) corresponds to
$p/\pz\ge0.4$ ($\le-0.4$ resp.). Time windows where $\max(|p/\pz|)<0.2$
are not taken into account in the data analysis (see text).
The gap lies between $t_s\simeq 13.5~\mu$s
and $t_r\simeq 15.0~\mu$s. The horizontal axis is the fast
``ultrasonic time'' $t$ and the vertical axis is the slow ``pulse time''
$T$. The rotor velocity is $\vr=23.5$~mm.s$^{-1}$.}
\end{figure}

Figure~\ref{f.data} shows twenty successive BS signals recorded in a sheared
Newtonian suspension with $\prf=1$~kHz. The ultrasonic
pulses leave the stator and enter the gap
at $t_s\simeq 13.5~\mu$s and a small fixed echo corresponding
to the rotor position may be seen at $t_r\simeq 15.0~\mu$s
(as well as a stronger echo
at about 15.2~$\mu$s that corresponds to a multiple reflection from
the Plexiglas window in the stator). 
The slopes of the traces left by the echoes in this
two-dimensional $(t,T)$ diagram are inversely proportional to the local
velocities according to Eq.~(\ref{e.vy}). Thus, the signature of
shear is rather clear: velocities increase from the stator, where
the scatterers remain almost fixed, to the rotor.
Moreover, the presence of moving echoes for times $t>t_r$ is not
surprising: these echoes simply correspond to
scattering of the wave reflected on the rotor.
One can see that the whole range of times from $t_s$ to $t_r$ gets
covered by some significant speckle signal
in about 20~ms in that case.

If better accuracy
is required, more averaging can be performed to improve the
statistical convergence of the method (see Fig.~\ref{f.converg} below).
Depending on the
trade-off between accuracy and temporal resolution,
we found that USV allows us to obtain a full velocity profile
every 0.02--2~s.

\section{Calibration procedure using a Newtonian fluid}
\label{s.calib}

\subsection{Newtonian suspension of ``contrast agents''}

In order to test and calibrate the experiment, much effort
was devoted to developing ``contrast agents'' that feature
good scattering properties for high-frequency USV.
Indeed, too weak a scatterer yields a signal
level too low to work with, whereas too strong a scatterer
leads to multiple scattering. When the scatterer
is small compared to the acoustic wavelength
(Rayleigh approximation), three parameters control
the amount of scattering: the scatterer diameter,
its compressibility and its density \cite{Pierce:1994}.
Moreover, if one wants the contrast agents to remain
Lagrangian tracers of the flow, their density has to
be matched to that of the fluid.

We found that homemade polystyrene spheres of diameter
$a=3$--10~$\mu$m were large and hard enough to scatter
36~MHz pulses efficiently but soft enough to prevent
multiple scattering when diluted at a weight concentration
of 1~\% in water. Such polystyrene spheres were obtained
by polymerization following Refs.~\cite{Echevarria:1998,Sugiura:2002}.
A 1~\% wt. solution of benzoyl peroxide in divynylbenzene is first
prepared. Then, a polydisperse emulsion is obtained by stirring 40~g
of the previous mixture in 59~g of water and 1~g of sodium
dodecyl sulfate at room temperature.
Finally, the emulsion is polymerized at 90$\degre$C for two hours. The prepared
polystyrene spheres were washed with water three times before drying. 
The 1~\% wt. suspension of such polystyrene spheres is
Newtonian and has the same viscosity as water.

\subsection{Measurement of $\cz$}

\begin{figure}[htbp]
\begin{center}
\scalebox{1}{\includegraphics{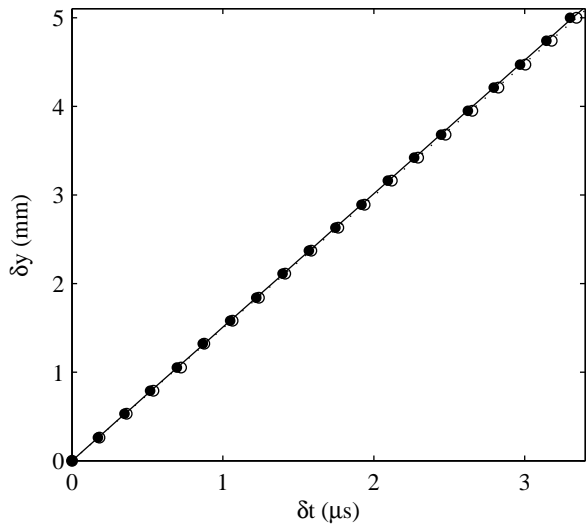}}
\end{center}
\caption{\label{f.soundspeed}Measurement of the sound speed $\cz$
as the slope of $\delta y$ vs. $\delta t$ (see text) for
the Newtonian 1~\% wt. suspension of polystyrene spheres
($\bullet$) and for the seeded lamellar phase of Sec.~\ref{s.lamellar}
($\circ$). Both data sets were obtained at 32$\degre$C and
$\delta y=0$ corresponds to a separation of about 10~mm between
the emitter and the receiver.
Linear fits yield $\cz=1507\pm 10$~m.s$^{-1}$ for the Newtonian suspension
(solid line) and $\cz=1495\pm 10$~m.s$^{-1}$ for the lamellar phase
(dotted line). In practice, those two fits are almost undistinguishable.
Typical errorbars on $\delta t$ are about the size of the symbols.}
\end{figure}

In order to use Eqs.~(\ref{e.vy}) and (\ref{e.yk}),
the sound speed $\cz$ of the
working fluid has to be measured. We used a classical
{\it transmission} setup
consisting of two transducers facing each other.
One of them, the emitter,
remains fixed while the other works as a receiver
and can be moved by a computer-controlled actuator.
Pulses are recorded for various displacements $\delta y$ of the receiver
and averaged over 100 sweeps. The time-shifts $\delta t$ between 
the arrival times of the pulses are measured
and the sound speed $\cz$ is given by the slope of $\delta y$ vs.
$\delta t$ (see Fig.~\ref{f.soundspeed}).

Due to technical constraints, the measurements of $\cz$ were
performed with 25~MHz transducers. Temperature
is controlled to within 0.1$\degre$C by a water
circulation around the sample. Figure~\ref{f.soundspeed}
shows that $\cz=1500$~m.s$^{-1}$ leads to a very good
approximation of the data obtained in the suspension of
polystyrene spheres at our working temperature of 32$\degre$C.
This value of $\cz$ is in good agreement with the measurements reported
in the literature for the sound speed of water vs. temperature
\cite{Pierce:1994}.
In the following, we will assume that this value does not vary
significantly with frequency between 25 and 36~MHz and all
the data will be taken at the same temperature of 32$\degre$C.

Finally, note that the measurement of $\cz$ using the transmission
of ultrasonic pulses between two transducers also allows us to check that
no significant multiple scattering takes place in our fluids by
looking at the transmitted waveform :
the received signal remains always as short as the emitted pulse
and no long-lasting echoes typical of multiply scattered waves
were ever recorded during the measurements of $\cz$.

\subsection{Measurement of $t_s$}

Another important parameter is $t_s$ that gives the position
of the stator--fluid interface and strongly depends on the
experimental conditions.

\begin{figure}[htbp]
\begin{center}
\scalebox{1}{\includegraphics{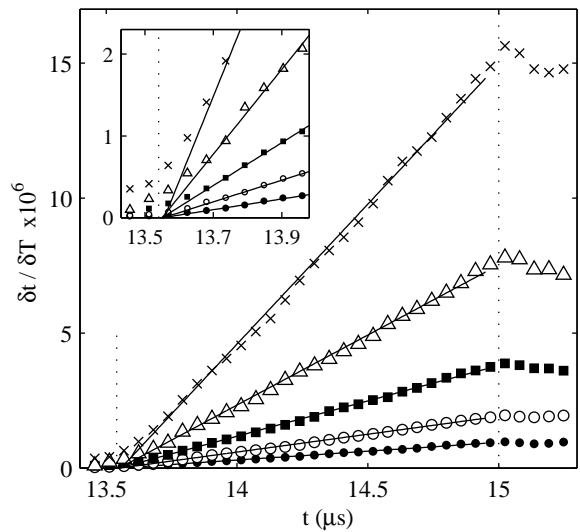}}
\end{center}
\caption{\label{f.brut}$\delta t/\delta T$
as a function of the position $t$ of the correlation window
obtained in the Newtonian suspension for
$\vr=2.9~(\bullet)$,
5.8~$(\circ)$, 11.7~$(\smblacksquare)$, 23.5~$(\smtriangle)$,
and 47.0~mm.s$^{-1}~(\times)$. The solid
lines are the best linear fits of the data for $13.55<t<14.95~\mu$s.
The dotted lines indicate the Plexiglas--fluid interfaces at the
stator (left) and at the rotor (right).
The results were averaged over 50 series of 20 pulses.
Typical standard deviations are of the order of
the marker size. Inset: blow-up of $\delta t/\delta T$ vs. $t$
close to the stator.}
\end{figure}

Figure~\ref{f.brut} presents the results of our speckle
tracking algorithm averaged over 50 series of 20 pulses
like that shown in Fig.~\ref{f.data}. The rotor velocity $\vr$
was varied between 2.9 and 47.0~mm.s$^{-1}$. The linear
behavior of the time-shift
$\delta t$ with the position $t$ of the correlation window
is a direct signature of the uniform shear flow inside the Newtonian
suspension. Indeed, all the velocities considered here are far
below the onset of the Taylor-Couette instability which occurs
when $\vr\,e^{3/2}/\nu\Rr^{1/2}\simeq 41$ (where
$\nu\simeq 10^{-6}$~m$^2$.s$^{-1}$ is the kinematic viscosity
of our calibration fluid)
\cite{Tritton:1988} {\it i.e.} when $\vr\simeq 200$~mm.s$^{-1}$.

For our dilute suspension, no wall slip is expected and $\delta t$
should go to zero at the stator: $\delta t(t_s)=0$. By linearly fitting
$\delta t/\delta T$ vs. $t$ for various values of $\vr$, different estimates of $t_s$ are
obtained that yield an average value $t_s=13.54\pm 0.02~\mu$s.
The uncertainty on $t_s$ is due to the presence of the stator
which leads to a small fixed echo on the signals and to small edge
effects around $t_s$ (see inset of Fig.~\ref{f.brut}).
It corresponds to an uncertainty of about $15~\mu$m on the position
of the stator--fluid interface.

\subsection{Conversion to orthoradial velocity profiles $v(x)$}
\label{s.conversion}

Inserting the values of $\cz$ and $t_s$ found above
in Eqs.~(\ref{e.vy}) and (\ref{e.yk}) yields the velocity profile $v_y(y)$.
Thus, so far, we obtained measurements of the velocity vector projected
along the acoustic axis. In order to get some physically more relevant
information, we will assume from now on that the radial velocity
is zero (or at least negligible when compared to the orthoradial velocity).
This is of course the case for our Newtonian suspension
below the onset of the Taylor-Couette instability. In complex
fluids, inhomogeneous flows with wall slip or shear banding
are purely orthoradial (see Sec.~\ref{ss.inhom})
so that the above approximation remains valid as long as
hydrodynamic or elastic instabilities do not occur.
Note that in the geometry discussed here, USV
measurements are not affected
by a non-zero vertical component of the velocity field.

\begin{figure}[htbp]
\begin{center}
\scalebox{1}{\includegraphics{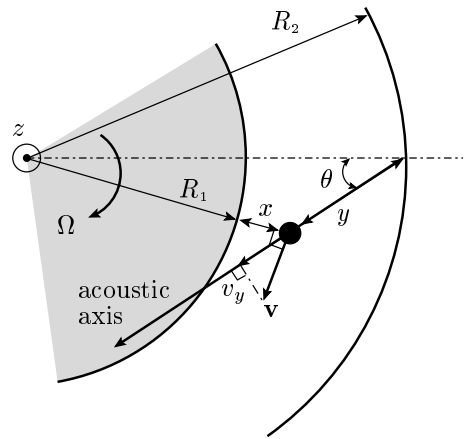}}
\end{center}
\caption{\label{f.convert}Conversion of $v_y(y)$ to $v(x)$ in the Couette
geometry. The flow is assumed
to be purely orthoradial. $x$ is the radial distance from the rotor whereas
$y$ is the distance from the stator--fluid interface along the acoustic axis.
The gap width was exaggerated to reveal curvature effects accounted
for by Eqs.~(\ref{e.v}) and (\ref{e.x}).}
\end{figure}

Assuming that $\mathbf{v}=(0,v,0)$ and using standard trigonometric
relations (see Fig.~\ref{f.convert}), one gets:
\begin{equation}
v(x) = \frac{\Rr+x}{\Rs\sin\theta}\, v_y(y)\,,
\label{e.v}
\end{equation}
where
\begin{equation}
x = \sqrt{\Rs^2+y^2-2\Rs y\cos\theta}-\Rr\,.
\label{e.x}
\end{equation}
In the small gap approximation,
Eqs.~(\ref{e.v}) and (\ref{e.x}) reduce to
$v(x)\simeq v_y/\sin\theta$ and $x\simeq e-y\sin\theta$.

Thus the last parameter to calibrate is the angle $\theta$.
This is done by looking for a value of $\theta$
for which $v(x)$ coincides with the velocity profile
expected for a Newtonian fluid:
\begin{equation}
v(x)=\vr\left(1+\frac{x}{\Rr}\right)\left[ 
\frac{ \left( \frac{\Rs}{\Rr+x} \right)^2 -1}{\left( \frac{\Rs}{\Rr} \right)^2 -1}\right]\simeq\vr\left(1-\frac{x}{e}\right)\,,
\label{e.vnewt}
\end{equation}
where the last term results from the small gap approximation.

\begin{figure}[htbp]
\begin{center}
\scalebox{1}{\includegraphics{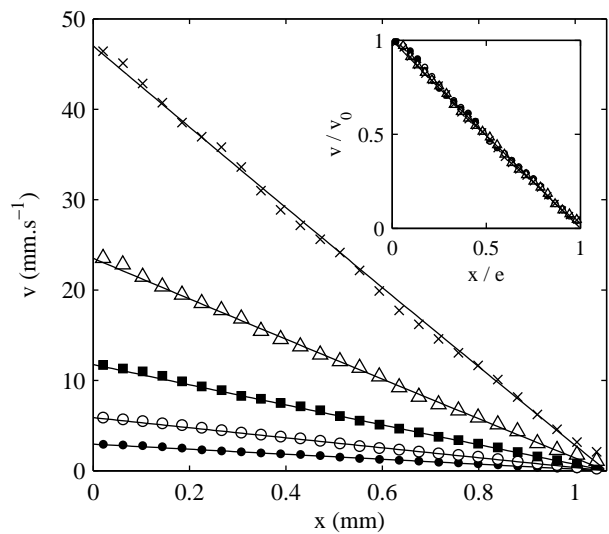}}
\end{center}
\caption{\label{f.calib}Velocity profiles $v(x)$ obtained from
Eqs.~(\ref{e.v}) and (\ref{e.x}) for the data of Fig.~\ref{f.brut}
with $\cz=1500$~m.s$^{-1}$, $t_s=13.54~\mu$s, and $\theta=13.5\degre$.
Rotor velocities are
$\vr=2.9~(\bullet)$, 5.8~$(\circ)$, 11.7~$(\smblacksquare)$, 
23.5~$(\smtriangle)$, and 47.0~mm.s$^{-1}~(\times)$.
The solid lines correspond to the Newtonian
velocity profiles given by Eq.~(\ref{e.vnewt}) with the above values
of $\vr$. Inset: dimensionless data $v/\vr$ vs. $x/e$.}
\end{figure}

Figure~\ref{f.calib} shows that $\theta=13.5\degre$
leads to very good results for all the rotor velocities $\vr$
considered here. Normalized data $v/\vr$ vs. $x/e$ show that
the Newtonian velocity profiles $v\simeq\vr(1-x/e)$ are recovered
for all our data with the set of parameters $\cz$, $t_s$, and $\theta$
inferred from the above calibration (see inset of Fig.~\ref{f.calib}).
This calibration method is very sensitive
to small variations of $\theta$ and allows its estimation to
within $0.2\degre$. Note that $\theta$ depends on both
the temperature and the nature of the fluid under study
through $\cz$.
As long as the temperature remains the same, the new value $\theta'$
of the incidence angle in a fluid whose sound speed $\cz'$
differs from that of the calibration fluid is simply given by 
Snell's law:
\begin{equation}
\sin\theta'=\frac{\cz'}{\cz}\sin\theta\,,
\end{equation}
where $\cz$ and $\theta$ are the parameters measured for the calibration
fluid.

\subsection{Statistical convergence of the USV measurements}

\begin{figure}[htbp]
\begin{center}
\scalebox{1}{\includegraphics{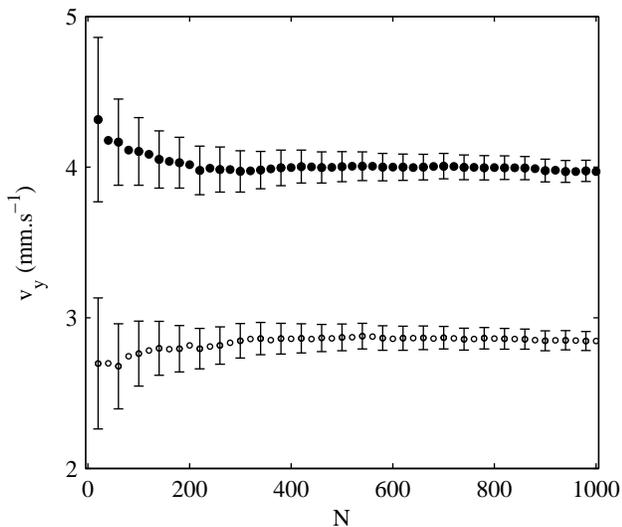}}
\end{center}
\caption{\label{f.converg}Convergence of the velocity estimates
as a function of the total number of pulses $N$ taken into account
in the analysis. $v_y$ is measured at
$y=0.6~(\circ)$ and 0.8~mm $(\bullet)$ for
$\vr=23.5$~mm.s$^{-1}$.}
\end{figure}

Finally,
Fig.~\ref{f.converg} shows the value of the $v_y$ measured at two
different positions along the acoustic beam as more and more averaging
is performed. It is clear that convergence is reached when
$N\simeq 300$ pulses
are taken into account. The standard deviation of the measurements
decreases from about 6~\% for $N\simeq 100$ to less than 3~\% for
$N>400$.

\section{Application to a complex fluid}
\label{s.lamellar}

\subsection{The lyotropic lamellar phase under study}

Let us now turn to an application of high-frequency USV to a
complex fluid. We chose to test USV in a lyotropic lamellar
phase composed of SDS (6.5~\% wt.), octanol (7.8~\% wt.) and brine
(20~g.L$^{-1}$ NaCl). This surfactant mixture is known to display
a series of structural shear-induced transitions \cite{Roux:1993,Diat:1995a,Sierro:1997}.
At equilibrium, the system is composed of surfactant bilayers
stacked with a smectic distance of about 15~nm. Under shear flow,
this lamellar phase was shown to form a closely packed assembly
of multilamellar vesicles
\cite{Diat:1995a}. The diameter of these vesicles, commonly called ``onions,''
is typically a few microns.

\begin{figure}[htbp]
\begin{center}
\scalebox{1}{\includegraphics{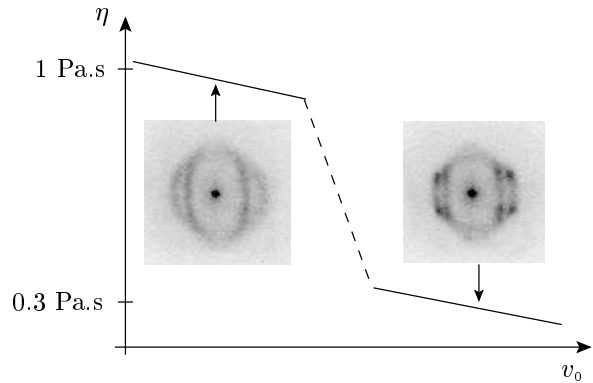}}
\end{center}
\caption{\label{f.transit}Typical evolution of the viscosity $\eta$
with the rotor velocity $\vr$ for the SDS/octanol/brine system.
A shear-thinning transition occurs as $\vr$ is increased
(dashed line).
The pictures show diffraction patterns observed in the low
shear regime (left) and on the high shear branch (right):
a laser beam crosses the cell along the $r$ direction and
a CCD camera collects scattered light in the plane $(\theta,z)$
\cite{Diat:1995a}.}
\end{figure}

At low shear velocities, the diffraction pattern 
observed with light scattering is an isotropic
ring indicating the absence of any positional order.
However, when the velocity $\vr$ is increased, Bragg peaks
appear in the diffraction pattern showing the apparition of long-range
order in the vesicle positions. As sketched in Fig.~\ref{f.transit},
this disorder-to-order transition corresponds to a sharp drop in the
sample viscosity by a factor of
about 3 \cite{Diat:1995a}. Such a shear-thinning transition is
also observed in colloidal suspensions and is called a ``layering''
transition \cite{Ackerson:1988}.
As explained in Sec.~\ref{ss.inhom}, inhomogeneous velocity profiles are
expected in the vicinity of this shear-induced transition where
the two states (ordered and disordered states) may coexist in the gap of a
Couette cell.

\begin{figure}[htbp]
\begin{center}
\scalebox{1.1}{\includegraphics{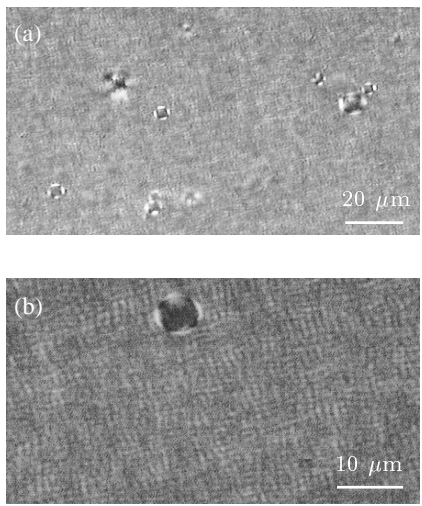}}
\end{center}
\caption{\label{f.photos}Sheared lamellar phase seeded with polystyrene
spheres of diameter 3--10~$\mu$m and observed under crossed polarizers at two different
magnifications: (a) $\times 20$ and (b) $\times 50$.}
\end{figure}

As for its ultrasonic behavior, the ``onion'' assembly
does not scatter 36~MHz pulses. Thus,
the lamellar phase was seeded with the polystyrene spheres used in
Sec.~\ref{s.calib} in order to get some speckle signal from the fluid.
We checked that the ``layering'' transition is not significantly
modified by the addition of such contrast agents. Figure~\ref{f.photos} 
presents two pictures of the seeded lamellar phase as seen under crossed
polarizers and after shearing:
the granular texture characteristic of disordered
multilamellar vesicles does not
seem perturbed by the addition of polystyrene spheres at 1~\%~wt.

\subsection{Evidence for inhomogeneous flows}

\begin{figure}[htbp]
\begin{center}
\scalebox{1}{\includegraphics{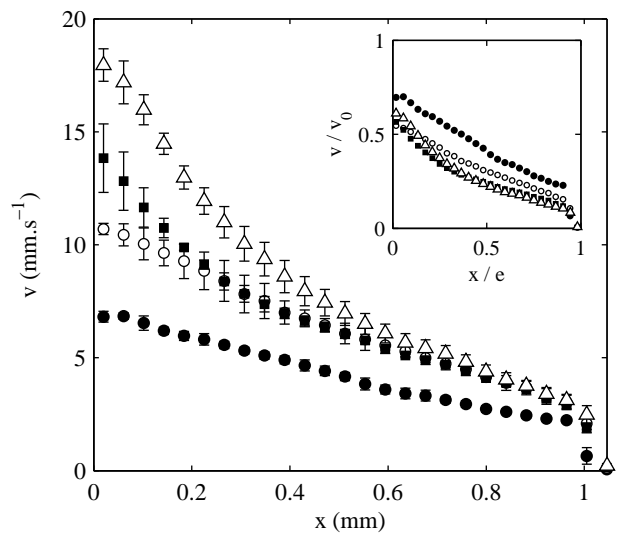}}
\end{center}
\caption{\label{f.lam1}Velocity profiles $v(x)$ 
measured in the lamellar phase for rotor velocities
$\vr=9.8~(\bullet)$, 19.6~$(\circ)$, 24.5~$(\smblacksquare)$,
and 29.4~mm.s$^{-1}~(\smtriangle)$. The data were
time averaged over several profiles (3 to 7 depending
on $\vr$) obtained from series of 1000 pulses recorded in about
2~s. Error bars correspond to the standard deviation of these
profiles and account mainly for temporal
fluctuations of the flow.
The inset shows the dimensionless data $v/\vr$
vs. $x/e$.}
\end{figure}

\begin{figure}[htbp]
\begin{center}
\scalebox{1}{\includegraphics{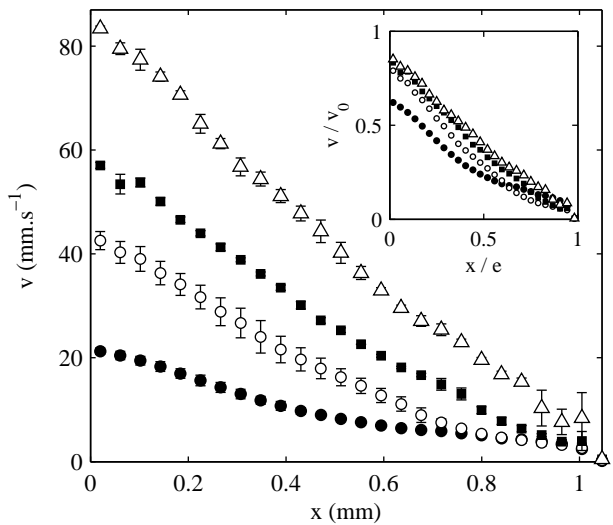}}
\end{center}
\caption{\label{f.lam2}Same as Fig.\ref{f.lam1} but with
$\vr=34.3~(\bullet)$, 53.8~$(\circ)$, 68.5~$(\smblacksquare)$,
and 97.9~mm.s$^{-1}~(\smtriangle)$.}
\end{figure}

As shown on Fig.~\ref{f.soundspeed}, the sound speed of the seeded
lamellar phase was measured to be $\cz=1495\pm 10$~m.s$^{-1}$.
Since the experiments and the calibration are performed at the
same temperature and since the sound speeds of both fluids are
the same (within less than 1~\%),
we used $\cz=1500$~m.s$^{-1}$, $t_s=13.54~\mu$s and $\theta=13.5\degre$ to
obtain the velocity profiles shown in Figs.~\ref{f.lam1} and
\ref{f.lam2}.

The two main features of the velocity profiles are as follows:

\paragraph{Wall slip.} As clearly shown by the insets of Figs.~\ref{f.lam1}
and \ref{f.lam2}, the velocity close to the walls is never equal to
that of the walls. On the contrary, $v(x=0)$ may differ
from $\vr$ by more than 30~\%. One can also notice that
slip velocities are comparable at the rotor and at the stator
for the smallest rotor velocity but become highly
dissymmetric when $\vr\gtrsim 20$~mm.s$^{-1}$. A detailed study
of wall slip in this system is left for future work.

\paragraph{Shear banding.} Figure \ref{f.lam1} also
unveils the presence of shear bands when $\vr\gtrsim20$~mm.s$^{-1}$.
A highly sheared band nucleates close to the rotor and
progressively fills the gap as $\vr$ is increased (see Fig.~\ref{f.lam2}
for the highest velocities). The shear rate in the high
shear band is about three times larger than that in the low shear band.
This ratio remains roughly constant as long as two bands coexist.
For $\vr\simeq 70$~mm.s$^{-1}$, the highly sheared region has invaded
the whole gap. When $\vr\gtrsim 70$~mm.s$^{-1}$, the flow is laminar
again and looks similar to that observed before the
``layering'' transition at $\vr\lesssim 20$~mm.s$^{-1}$.

Simultaneously to the present
ultrasonic measurements, DLS data obtained in our group
have confirmed the existence of shear-banded flows and strong
wall slip effects in the same lamellar phase without seeding 
polystyrene particles \cite{Salmon:2003d}. This last observation
shows {\it a posteriori} that our ``contrast agents''
do not perturb the flow and can be treated as simple Lagrangian tracers.
Both USV and DLS data are in good qualitative agreement
with classical theoretical pictures
of shear-banded flows \cite{Spenley:1993,Olmsted:1997}.
Our purpose here is only to show the relevance of high-frequency USV
in the field of sheared complex fluids and a more quantitative
analysis of shear banding in light of DLS results
can be found in Ref.~\cite{Salmon:2003d}.

\subsection{Dynamical measurements during a transient}

Finally let us show how USV may be used to follow the dynamics
of a complex fluid under shear. Until now, we have considered
time-averaged velocity profiles. However, during the ``layering''
transition, noticeable fluctuations of the flow field occur and
result in large error bars for $\vr=24.5$ or 29.4~mm.s$^{-1}$ for
instance. Actually, some very rich dynamics were reported
in this sheared lamellar phase
when the {\it torque} applied on the rotor is imposed instead
of the {\it velocity} $\vr$ \cite{Wunenburger:2001}.
In that case, the viscosity $\eta(t)$ displays
large, complex fluctuations on long time scales of the order
of 100~s. Similar (although faster) temporal behaviors
were also observed in other surfactant systems
\cite{Wheeler:1998,Bandyopadhyay:2001}.

The origin of these fluctuations and their precise statistical properties
remain unclear.
One possibility is that the flow presents shear
bands that may couple to each other and lead to the observed behavior of
the viscosity. Thus the dynamics could be not only temporal as observed
on $\eta(t)$ but also spatial as suggested in Ref.~\cite{Salmon:2002}.

\begin{figure}[htbp]
\begin{center}
\scalebox{1}{\includegraphics{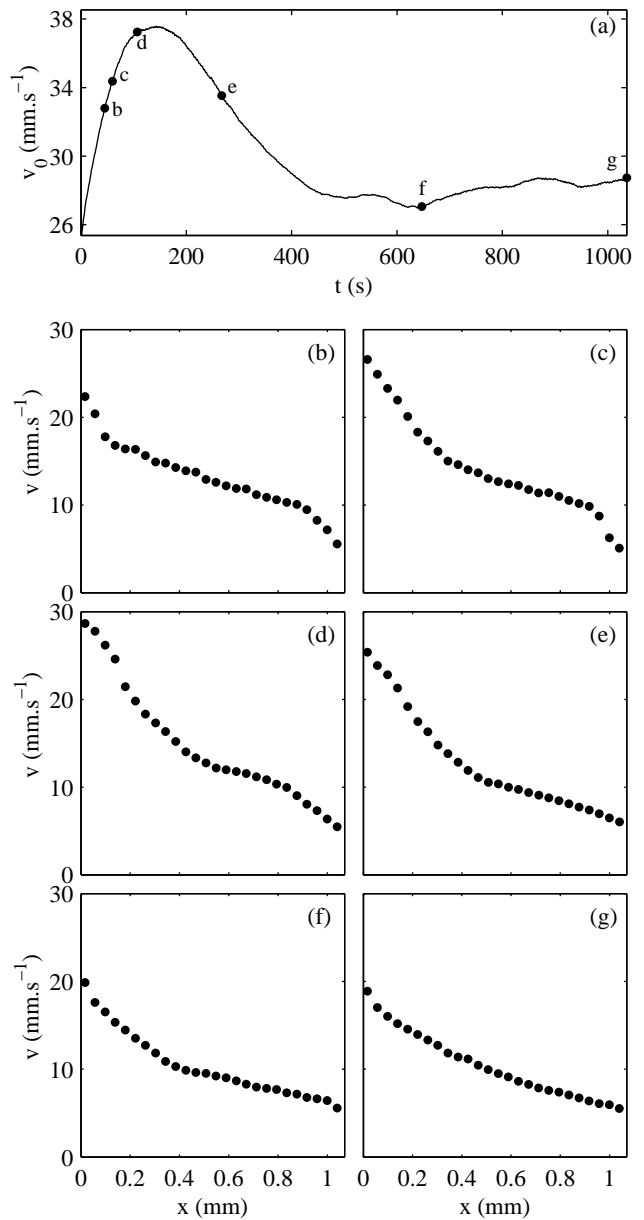}}
\end{center}
\caption{\label{f.lam3}Transient regime observed when the torque
is increased from 83 to 87~mN at $t=0$. (a) Velocity $\vr$ of the rotor
as a function of time $t$. Bullets ($\bullet$) correspond
to the points where the velocity profiles shown in (b)--(g) were measured:
(b) $t=45$~s, (c) $t=60$~s, (d) $t=107$~s,
(e) $t=267$~s, (f) $t=647$~s, and (g) $t=1036$~s.
The velocity profiles were
obtained from series of 800 pulses recorded in 1.6~s.}
\end{figure}

Figure~\ref{f.lam3} shows a transient regime recorded after
the torque applied to
the rotor was suddenly increased from 83 to 87~mN at time $t=0$.
The response of the rotor velocity $\vr(t)$ is shown
in Fig.~\ref{f.lam3}(a) \cite{Remark:moviesUSV}. A maximum
of about 38~mm.s$^{-1}$ is reached at $t\simeq 300$~s before
$\vr(t)$ slowly relaxes with small fluctuations.
Velocity profiles recorded during that transient show
that: (i) during the initial increase of $\vr(t)$,
a shear-banded flow is generated with
{\it three} bands: two highly sheared bands at both the
rotor and the stator and a low shear region in between
(see Fig.~\ref{f.lam3}(b)--(d)); (ii) the band located
at the stator disappears when $\vr(t)$ decreases, leaving
only {\it two} bands similar to Figs.~\ref{f.lam1}
and \ref{f.lam2} (see Fig.~\ref{f.lam3}(e)--(f)); (iii)
states where the highly sheared band almost disappears
can be observed (see Fig.~\ref{f.lam3}(g)) even though
the rotor velocity $\vr\gtrsim 28$~mm.s$^{-1}$
indicates that a band should extend over almost half the gap 
if the average picture at imposed
$\vr$ remained valid when the torque is imposed
(see the $\smtriangle$ symbols corresponding
to $\vr=29.4$~mm.s$^{-1}$ in Fig.~\ref{f.lam1}).

\section{Discussion and conclusions}
\label{s.concl}

\subsection{Summary}

The above results show that high-frequency USV is a powerful tool
for measuring velocity profiles in sheared complex fluids.
We have successfully tested the technique on a lyotropic
lamellar phase that undergoes a shear-induced ``layering''
transition.

When the rotor velocity, {\it i.e.} the global
shear rate, is imposed, shear banding is observed: 
time-averaged velocity profiles display a highly sheared
region that nucleates from the rotor and progressively invades the
gap as $\vr$ is increased. Wall slip can also be very large
especially during the transition, which makes global measurements
such as the sample viscosity tricky to analyze.

Moreover, USV allowed us to record velocity profiles in 
1~s typically. This acquisition time is short enough (compared
to the intrinsic time scales of our complex fluid)
to follow the velocity profiles in time. We showed the
existence of more complex, three-band flows during a transient
at imposed torque {\it i.e.} imposed shear stress.
This raises the question of the structure of the flow during
the spontaneous oscillations of the viscosity observed in
the same lamellar phase \cite{Wunenburger:2001,Salmon:2002}.
USV may thus provide a
``time-resolved'' tool for investigating spatially the dynamical
behaviors of $\eta(t)$.

\subsection{Comparison with other local techniques}

In this paper, we have shown that we were able to measure
velocity profiles with a spatial resolution of about
$40~\mu$m and every 0.02--2~s. High-frequency USV compares
very well with other local
techniques for measuring flow fields.

LDV, or equivalently
DLS, provides in principle a slightly better spatial resolution
of 10--50~$\mu$m but only yields a pointlike measurement.
In order to obtain a full velocity profile with LDV or DLS,
one has to mechanically scan the whole gap of the cell
\cite{Salmon:2003b}.
This usually takes about 1~min and may raise major interpretation
problems when the flow evolves on time scales shorter than 1~min.

NMR appears as a very promising technique for complex fluids
\cite{Hanlon:1998,Mair:1996,Britton:1999}
since no seeding is required and two-dimensional 
images of the flow are obtained. However, acquisition
times are rather long too (from about 20~s to more than
an hour depending on the number of points in the image),
making it rather difficult to access dynamical local information.

The main drawback of USV is that it needs a good control of
the scattering properties of the fluid and that it may
require that the flow be seeded with contrast agents. Unless
their mechanical properties is strictly controlled, most
concentrated suspensions or emulsions will lead to multiple
scattering of high-frequency pulses. Thus, further application
of USV to other complex fluids will involve a careful study
of their acoustic properties. However, we believe that USV
is likely to extend the range of complex fluids in which local
measurements may be performed, in particular to optically
opaque fluids.

\subsection{Perspectives}

The biggest advantage
of USV lies undoubtedly in its ability to perform
nonintrusive velocity measurements fast enough to follow the dynamics
of complex fluids under low shear. In principle, in an ideal scattering
environment, the temporal resolution of USV could even get as short as
1~ms per measurement.

Future research directions on USV are as follows: (i) more
measurements on seeded lamellar phases will be performed
and compared to DLS data;
more precisely, we will focus on
the spatio-temporal flow dynamics in regimes where sustained
oscillations of the viscosity are observed; (ii) we will investigate
shear flows of other fluids such as emulsions, gels, or granular pastes;
(iii) we also intend to apply high-frequency USV to the case
of very small deformations in order to perform some ``local
linear rheology'' and understand how a complex fluid starts to flow
in the transition from linear to nonlinear rheology.

\begin{acknowledgments}
The authors wish to thank the ``Cellule Instrumentation'' at CRPP
for designing and building the mechanical parts of the experimental setup.
We are very grateful to D. Roux, J.-B. Salmon, and R. Wunenburger
for fruitful discussions.
\end{acknowledgments}

\end{document}